\documentclass{article}
\usepackage{spconf,amsmath,graphicx}

\def\given{\:|\:}
\def\L{\mathsf{L}}
\def\H{\mathsf{H}}

\newcommand{\ciid}{C_{\mathrm{iid}}}

\newcommand{\xlevel}{\mathsf{x}}
\newcommand{\xl}{\xlevel_\L}
\newcommand{\xh}{\xlevel_\H}

\title{Finite-State Channel Models for Signal Transduction in Neural Systems}
\twoauthors
  {Andrew W. Eckford\sthanks{$^*$Corresponding author. Email: aeckford@yorku.ca}\thanks{AWE was funded in part by a grant from the Natural Sciences and Engineering Research Council (NSERC).}}
	{Dept. of EECS\\
	York University\\
	Toronto, Ontario, Canada M3J 1P3}
  {Kenneth A. Loparo$^1$ and Peter J. Thomas$^2$\thanks{PJT was supported by NSF grants EF-1038677 and DMS-0720142.}}
	{$^1$Dept. of EECS and $^2$Dept. of Mathematics\\
	Case Western Reserve University\\
	Cleveland, Ohio, USA 44106-7058}
\begin{document}
%
\maketitle
\begin{abstract}
Information theory provides powerful tools for understanding communication systems.
This analysis can be applied to intercellular signal transduction, which is a means of 
chemical communication among cells and microbes.
We discuss how to apply information-theoretic analysis to 
ligand-receptor systems, which form the signal carrier and receiver in intercellular signal transduction
channels. We also discuss the applications of these results to neuroscience.
\end{abstract}

\section{Introduction}

The human brain is a vast communications engine, comprising some 100 billion nerve cells connected by upwards of 100 trillion synapses.
Information theory has a long history of application in the biological sciences generally \cite{yockey1958a} and neuroscience in particular \cite{mackay1952}.  Capacity and mutual information have proven fruitful concepts in understanding sensory systems \cite{Barlow1961,BellSejnowski1997VisRes,OlshausenField2004CurrOpNeurobio},  fault tolerant computation \cite{kn:CowanWinograd};  computation in spiking neurons \cite{Ruyter-van-SteveninckLewenStrongKoberleBialek1997Science,ToupsFellousThomasSejnowskiTiesinga2012PLOS_CB}; biological computation under metabolic constraints \cite{Laughlin2001CurrOpinNeuro,BergerLevy2010IEEE_IT,XingBergerSejnowski2012ISIT}, and information processing limitations of  genetic regulatory elements \cite{TkavcikCallanBialek2008PRE,TkavcikCallanBialek2008PNAS}.

Here we discuss some capacity bounds for several signaling  systems present in the brain.  A common motif in neurobiological communication is the transduction of  chemical, mechanical, or optical signals into ionic currents across the membrane of a nerve cell.  
Signal transduction typically involves specialized protein molecules: rhodopsin can detect single photon absorptions in the retina \cite{RiekeBaylor1998RMP}; 
acetylcholine receptor proteins convert the chemical neurotransmitter signal into muscle-activating currents to move the limbs \cite{FambroughDrachmanSatyamurti1973Science}.   
We focus here on two examples: channelrhodopsin (ChR, widely used as a control mechanism for neuroscience experiments) and the acetylcholine receptor (AChR).  Both systems convert their signals into an all-or-none conductance, effectively acting as graded input, binary output systems; both have multiple internal states (three for ChR, five for AChR).  Hence the state of the system is only \emph{partially observable}, complicating capacity estimates.


\section{Model}

\subsection{Master equation kinetics}
\label{sec:MasterEquation}

For a receptor with $k$ discrete states, 
there exists a $k$-dimensional vector of state occupancy probabilities $p(t)$,
given by
\begin{equation}
	p(t) = \left[ p_1(t), \: p_2(t), \: \ldots, \:p_{k}(t) \right] ,
\end{equation}
where $p_i(t)$ represents the probability of a given receptor occupying state $i$ at time $t$.
The chemical kinetics of the receptor are captured by a differential equation known
as the {\em master equation}. Let $Q = [q_{ij}]$ represent a $k \times k$ matrix of rate constants,
where $q_{ij}$ represents the instantaneous rate at which receptors starting in state $i$
enter state $j$. Then the master equation is given by
%
%
$dp/dt=p(t)Q$.

We use the notation from \cite{GroffDeRemigioSmith2009chapter}.
In the following examples:
\begin{itemize}
	\item Rates which are sensitive to the input are directly proportional to the input $x(t)$: for example, 
	$q_{12}$ is the transition rate from state 1 to state 2, 
	which is not sensitive to the input; while 
	$q_{30}x(t)$ is the transition rate from state 3 to
	state 0, sensitive to the input; and
	\item The $i$th diagonal element is written $R_i$, and is set so that the $i$th row sums to zero (so, if $x(t)$ appears in the $i$th row, $R_i$ depends on $x(t)$).
\end{itemize}

\begin{figure}
	\begin{center}
	\includegraphics[width=1.4in]{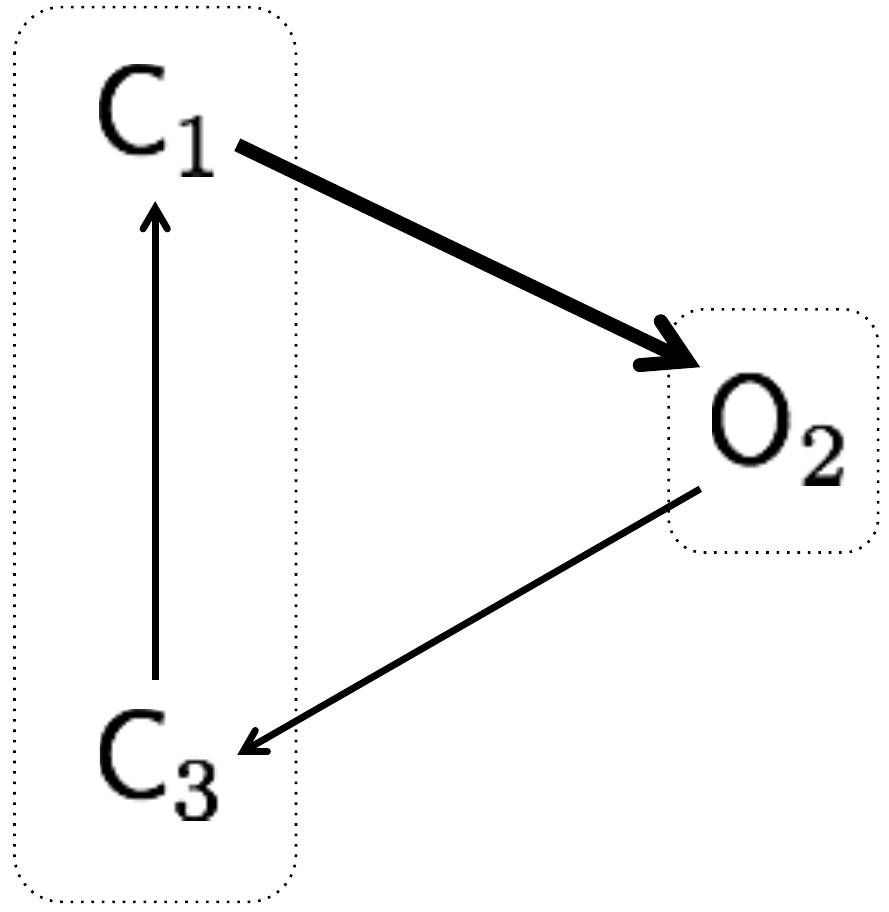}
	\end{center}
	\caption{\label{fig:ChR2} Depiction of allowed state transitions for ChR2. Sensitive 
	transitions are depicted with {\bfseries bold} arrows. 
	States are labelled by their ion channel state: $\{\mathsf{C},\mathsf{O}\}$ for closed and open, respectively; state number is in subscript. Dashed lines surround all states in either the closed or open state. Transition rates, listed in Table \ref{tab:ChR2parameters}, correspond to the vertices associated with each directed edge: for example,
	the rate from state $\mathsf{O}_2$ to state $\mathsf{C}_3$ is $q_{23}$.}
\end{figure}

{\em Example 1: Channelrhodopsin-2 (ChR2)}. 
The ChR2 receptor is a light-gated ion channel.
The receptor has three states, named Closed ($\mathsf{C}_1$), Open ($\mathsf{O}_2$), and Desensitized ($\mathsf{C}_3$).
The channel-open ($\mathsf{O}$) state $\mathsf{O}_2$ is the only state in which the ion channel is open, passing an ion current.
The channel-closed ($\mathsf{C}$) states, $\mathsf{C}_1$ and $\mathsf{C}_3$, are distinct in that the receptor is light-sensitive in
state $\mathsf{C}_1$, and insensitive in state $\mathsf{C}_2$ \cite{nag03}. The rate matrix for ChR2 is
\begin{equation}
	Q = \left[
		\begin{array}{ccc}
			R_1 & q_{12}x(t) & 0 \\
			0 & R_2 & q_{23} \\
			q_{31} & 0 & R_3
		\end{array}
	\right] .
\end{equation}
Fig.~\ref{fig:ChR2} shows
state labels and allowed state transitions.
Parameter values from the literature are given in Table \ref{tab:ChR2parameters}. We will assume,
as in \cite{tch13}, that the opening rate $q_{12}x(t)$ is directly proportional to the 
irradiance of the light on the receptor.
\begin{table}[h!]\begin{center}
	\begin{tabular}{c|c|c} \hline
		Parameter & from \cite{nag03} & Units \\ \hline
		$q_{12}x(t)$ & $\leq 5 \times 10^3$ &  s$^{-1}$   \\ \hline
		$q_{23}$ & 50 &  s$^{-1}$   \\ \hline
		$q_{31}$ & 17 &  s$^{-1}$ \\ \hline
	\end{tabular}
	\ \\
	\caption{\label{tab:ChR2parameters} Rate parameters for ChR2, adapted from \cite{nag03}.}
	\end{center}
\end{table}

{\em Example 2: Acetylcholine (ACh)}.
The ACh receptor is a ligand-gated ion channel. The receptor has five states, with rate matrix
\begin{equation}
	\label{eqn:AChRateMatrix}
	Q =
	\left[
		\begin{array}{ccccc}
			R_1 & q_{12}x(t) & 0 & q_{14} & 0 \\
			q_{21} & R_2 & q_{23} & 0 & 0 \\
			0 & q_{32} & R_3 & q_{34} & 0 \\
			q_{41} & 0 & q_{43}x(t) & R_4 & q_{45} \\
			0 & 0 & 0 & q_{54}x(t) & R_5 
		\end{array}
	\right] .
\end{equation}
%
%

\begin{figure}
	\begin{center}
	\includegraphics[width=2in]{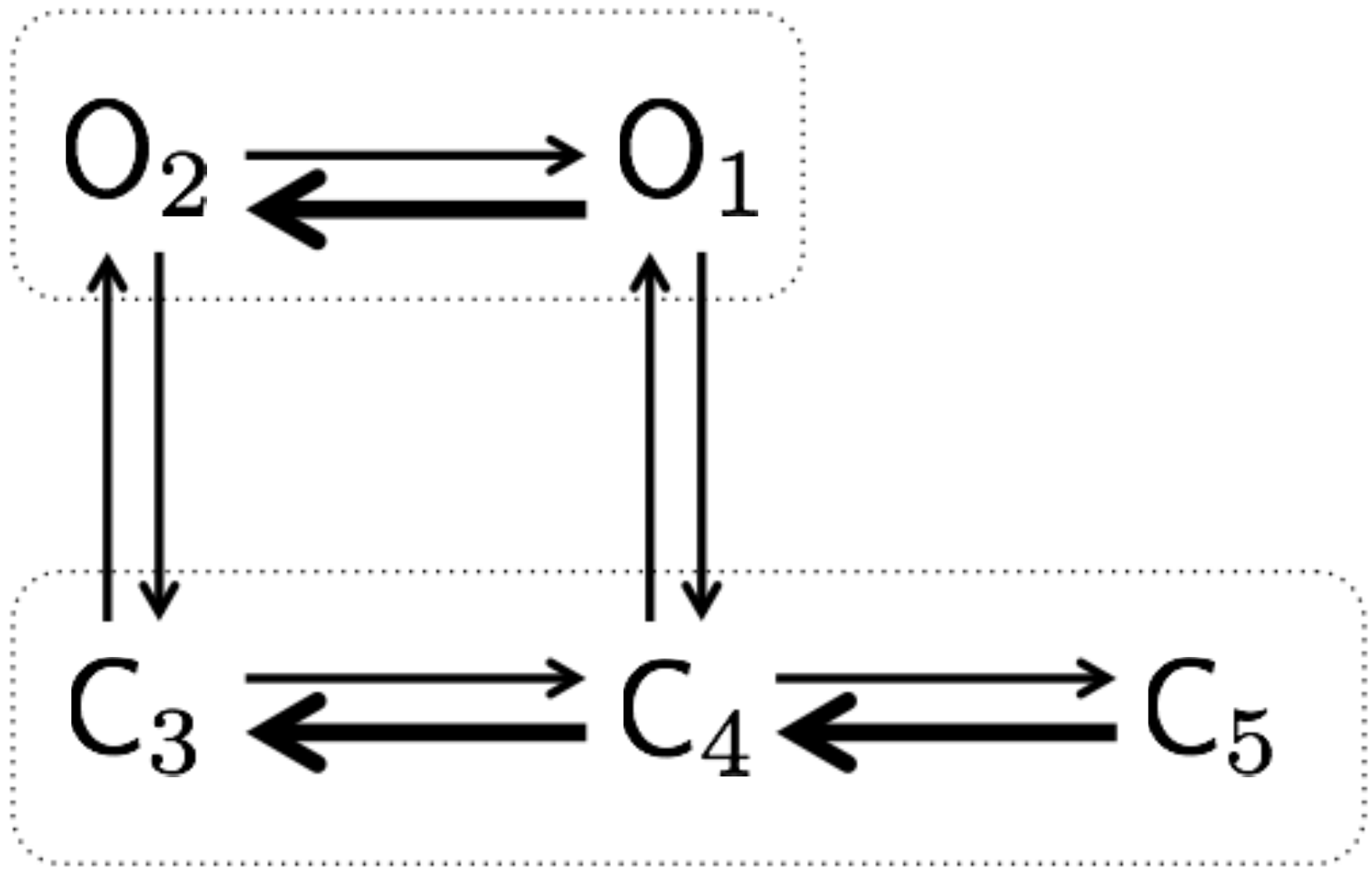}
	\end{center}
	\caption{\label{fig:ACh} Depiction of allowed state transitions for ACh. Sensitive 
	transitions are depicted with {\bfseries bold} arrows. 
	States are labelled by their ion channel state: $\{\mathsf{C},\mathsf{O}\}$ for closed and open, respectively; state number is in subscript. Dashed lines surround all states in either the closed or open state. Transition rates, listed in Table \ref{tab:AChparameters}, correspond to the vertices associated with each directed edge: for example,
	the rate from state $\mathsf{O}_2$ to state $\mathsf{C}_3$ is $q_{23}$.}
\end{figure}

There are three sensitive transitions: $r_{12}x(t)$, $r_{43}x(t)$, and $r_{54}x(t)$, which 
are proportional to agonist concentration $x(t)$. These transitions represent binding of an ACh molecule to one of two binding sites.
Fig.~\ref{fig:ACh} shows the
allowed state transitions. State $\mathsf{C}_5$ corresponds to both sites unoccupied; states $\mathsf{C}_4$, $\mathsf{O}_1$ correspond to one site occupied; states $\mathsf{C}_3$, $\mathsf{O}_2$ correspond to both sites occupied.  Table \ref{tab:AChparameters} gives
parameter values; the concentration of ACh, $x(t)$, is measured in mol/$\ell$.

The same state-naming convention is used in the figure as with ChR2: states with
an open ion channel are $\mathsf{O}_{1}$ and $\mathsf{O}_2$; states 
with a closed ion channel are $\mathsf{C}_3$, $\mathsf{C}_4$, and $\mathsf{C}_5$.

\begin{table}
	\begin{tabular}{c|c|c|c} \hline
		Parameter & Name in \cite{col82} & Value/range & Units \\ \hline
		$q_{12}x(t)$ & $k_{+2}x$ & $5 \times 10^8 x(t)$ & s$^{-1}$ \\ \hline
		$q_{14}$ & $\alpha_1$ & $3 \times 10^3$ & s$^{-1}$ \\ \hline
		$q_{21}$ & $2 k_{-2}^*$ & $ 0.66 $ & s$^{-1}$ \\ \hline
		$q_{23}$ & $\alpha_2$ & $5 \times 10^2$ & s$^{-1}$ \\ \hline
		$q_{32}$ & $\beta_2$ & $1.5 \times 10^4$ & s$^{-1}$ \\ \hline
		$q_{34}$ & $2 k_{-2}$ & $ 4 \times 10^3$ & s$^{-1}$ \\ \hline 
		$q_{41}$ & $\beta_1$ & 15 & s$^{-1}$ \\ \hline
		$q_{43}x(t)$ & $k_{+2}x$ & $(5 \times 10^8) x(t)$ & s$^{-1}$  \\ \hline
		$q_{45}$ & $k_{-1}$ & $ 2 \times 10^3$ & s$^{-1}$ \\ \hline
		$q_{54}x(t)$ & $2 k_{+1} x$ & $(1 \times 10^8) x(t)$ & s$^{-1}$  \\ \hline
	\end{tabular}
	\ \\

	\caption{\label{tab:AChparameters}Rate parameters for ACh, adapted from \cite{col82}, where $x(t)$ represents the molar concentration of ACh in mol/$\ell$.}
\end{table}

For each of the preceding examples, the rate constants depend on environmental conditions, and
thus can be reported differently in different sources 
(e.g., \cite{lin09} for ChR2).

\subsection{From the master equation to discrete-time Markov chains}

It is possible to discretize the master equation and describe the dynamics of a 
receptor as a discrete-time Markov chain; this is important to our paper as we rely on
capacity results for discrete-time Markov channels. Briefly,
we can discretize the master equation by writing
\begin{equation}
	\label{eqn:Markov-1}
	p(t + \Delta t)  = p(t) \left( I + \Delta t Q \right)+o(\Delta t)
\end{equation}
where $I$ is the identity matrix, and $o(\Delta t)/\Delta t\to 0$ as $\Delta t\to 0$.
If we let 
%
	$p[j] = p(j \Delta t)$,
%
then (\ref{eqn:Markov-1}) becomes
%
	$p[j+1] = p[j] (I+\Delta t Q)$.
%
Thus, we have a discrete-time Markov chain with transition probability matrix 
\begin{equation}
	\label{eqn:Markov-last}
	P = I + \Delta t Q.
\end{equation}
The matrix $P$ satisfies the conditions of a Markov chain transition
probability matrix (nonnegative, row-stochastic) as long as $\Delta t$ is small enough.  Note that while the probability $p(t)$ evolves deterministically, the channel state itself is a non-Gaussian random process taking discrete values.  

\section{Signal transduction as a communications system}

\subsection{Communication model of receptors}

We now discuss how the receptors can be described as 
information-theoretic communication systems: that is, in terms of input, output, and conditional
input-output PMF.

{\em Input:} The receptor input $x(t)$ consist of either light intensities or ligand concentrations, and is discretized in time: for integers $i$, the input is $x(i \Delta t)$; we will write
$x_i = x(i \Delta t)$. 
We will also discretize the amplitude,
so that for every $t$, $x_i \in \{\xlevel_1,\xlevel_2,\xlevel_3,\ldots,\xlevel_k\} =: \mathcal{X}$. 
We will assume that the $\xlevel_i$ are 
distinct and increasing; further, we assign the lowest value $\xlevel_1$ and the highest value $\xlevel_k$
the symbols $\xl$ and $\xh$, respectively.

{\em Output:} 
Receptor states from our example systems are labelled, e.g., $\mathsf{C}_3$ or $\mathsf{O}_2$.
The output of the communication system is given either by: 
the receptor state $y(t)$, given by the {\em subscript} of the state label; or by the ion channel state
$z(t)$, either $\mathsf{C}$ or $\mathsf{O}$, without subscript.
These are discretized, respectively, to $y_i = y(i\Delta t)$ and $z_i = z(i \Delta t)$.

{\em Conditional input-output PMF:} From (\ref{eqn:Markov-1})-(\ref{eqn:Markov-last}), $y^n$ forms a 
Markov chain given $x^n$, so 
\begin{equation}
	\label{eqn:ReceptorMarkov}
	p_{Y^n|X^n}(y^n|x^n) = \prod_{i=1}^n p_{Y_i \given Y_{i-1},X_i}(y_i \given y_{i-1},x_i) ,
\end{equation}
where $p_{Y_i \given Y_{i-1},X_i}(y_i \given y_{i-1},x_i)$ is given by the appropriate entry in the matrix $P$, and where $y_0$ is null.%
\footnote{We say a variable is ``null'' if it vanishes under conditioning, i.e., if $y_0$ is null, then $p_{Y_1 | X_1, Y_0}(y_1 \given x_1, y_0) = p_{Y_1 | X_1}(y_1 \given x_1)$.}
For example, using ACh, suppose $y_{i-1} = 1$, $y_i = 2$, and $x_i = \xh$. Then from 
(\ref{eqn:Markov-last}) and Table \ref{tab:AChparameters}, we have
$p_{Y_i \given Y_{i-1},X_i}(2 \given 1,\xh) = \Delta t q_{12}(t) = 5 \times 10^8 \xh \Delta t$.

\subsection{Information theory and Shannon capacity}

We briefly review the information-theoretic concepts used in the paper. The reader is
directed to \cite{cover-book} for further detail.

A communication channel consists of: a vector of inputs $[x_1,x_2,\ldots]$, a vector of outputs 
$[y_1,y_2,\ldots]$,
and a conditional 
probability density function relating outputs to inputs.
Using the following vector notation: 
\begin{align} x^n &= [x_1,x_2,\ldots,x_n] \\ y^n &= [y_1,y_2,\ldots,y_n] , \end{align}
the stochastic input-output relationship is given by the 
conditional joint
PMF $p_{Y^n|X^n}(y^n|x^n)$.
 
For a channel with inputs $x^n$ and outputs $y^n$, 
the mutual information $I(X^n;Y^n)$ gives the maximum information rate that may be 
transmitted reliably over the channel.
Mutual information is given by
\begin{align}
	\label{eqn:MutualInformation0}
	I(X^n;Y^n) &= E \left[ \log \frac{p_{Y^n|X^n}(y^n\given x^n)}{p_{Y^n}(y^n)} \right] \\
\end{align}
where $p_{Y^n |X^n}(y^n \given x^n)$ is the conditional probability mass function (PMF) of $Y^n$ 
given $X^n$,
and $Y$,
and $p_Y(y)$ is the marginal PMF on $Y$.
%

As $n \rightarrow \infty$, generally $I(X^n;Y^n) \rightarrow \infty$ as well; 
in this case, it is useful to calculate the mutual information rate, given by
\begin{align}
	\nonumber
	\lefteqn{\mathcal{I}(X;Y)} & \\
	\label{eqn:InformationRate}
	&= \lim_{n \rightarrow \infty} \frac{1}{n} \sum_{x^n,y^n} 
		p_{X^n,Y^n}(x^n,y^n) 
		\log \frac{p_{Y^n|X^n}(y^n \given x^n)}{p_{Y^n}(y^n)} .
\end{align}
We will assume that receptor response is stationary.
Similar derivations hold for $\mathcal{I}(X;Z)$, the mutual information rate from inputs
to ion channel state.

\subsection{Receptor IID Capacity}

The capacity $C$ of a communication system is the maximum over all possible 
input distributions $p_{X^n}(x^n)$ of $\mathcal{I}(X;Y)$.
If the inputs $p_{X^n}(x^n)$ is restricted to the set of 
independent, identically distributed (IID) input distributions,
i.e. we can write $p_{X^n}(x^n) = \prod_{i=1}^n p(x_i)$,
then we have the IID capacity, written $\ciid$. It should be clear that $\ciid < C$.

We now calculate $\ciid$ 
for the discrete-time receptor model. Although
IID inputs are not realistic in practice (as concentration may persist for long periods of time),
they can be capacity-achieving under some circumstances \cite{eck13}. 

In general, since $Y^n$ is a time-inhomogeneous Markov chain if $X^n$ is known,
we can write
\begin{equation}
	\label{eqn:Conditional-1}
	p_{Y^n | X^n}(y^n \given x^n) = \prod_{i=1}^n p_{Y_i | X_i, Y_{i-1}}(y_i \given x_i, y_{i-1}),
\end{equation}
Under IID inputs, it can be shown that the receptor states $Y^n$ form a time-homogeneous Markov
chain, that is,
\begin{equation}
	\label{eqn:Conditional-2}
	p_{Y^n}(y^n) = \prod_{i=1}^n p_{Y_i | Y_{i-1}}(y_i \given y_{i-1}),
\end{equation}
where $y_0$ is again null, and where 
\begin{equation}
	p_{Y_i | Y_{i-1}}(y_i \given y_{i-1}) = \sum_x p_{Y_i | X_i, Y_{i-1}}(y_i \given x, y_{i-1}) p_X(x) .
\end{equation}
Using (\ref{eqn:Conditional-1})-(\ref{eqn:Conditional-2}), (\ref{eqn:MutualInformation0})
reduces to
\begin{equation}
	I(X^n;Y^n) = \sum_{i=1}^n E\left[ \log \frac{p_{Y_i \given X_i, Y_{i-1}}(y_i \given x_i,y_{i-1})}
	{ p_{Y_i | Y_{i-1}}(y_i \given y_{i-1})}\right]
\end{equation}
and (\ref{eqn:InformationRate}) reduces to
\begin{equation}
	\label{eqn:MarkovInformationRate}
	\mathcal{I}(X;Y) = E\left[ \log \frac{p_{Y_i \given X_i, Y_{i-1}}(y_i \given x_i,y_{i-1})}
	{ p_{Y_i | Y_{i-1}}(y_i \given y_{i-1})}\right]
\end{equation}

Considering the diagrams in the previous section, some of the transitions 
were sensitive (i.e., dependent on input $x_i$), and others were insensitive (i.e.,
independent of $x_i$). From (\ref{eqn:MarkovInformationRate}), if the transition
$p_{Y_i \given X_i, Y_{i-1}}(y_i \given x_i,y_{i-1})$ is insensitive, then
\begin{align}
	\log \frac{p_{Y_i | X_i, Y_{i-1}}(y_i \given x_i,y_{i-1})}
	{ p_{Y_i | Y_{i-1}}(y_i \given y_{i-1})}
		&= \log \frac{p_{Y_i | Y_{i-1}}(y_i \given y_{i-1})}
	{ p_{Y_i | Y_{i-1}}(y_i \given y_{i-1})} \\
		&= \log 1 = 0.
\end{align}
Thus, (\ref{eqn:MarkovInformationRate}) is calculated using the {\em sensitive transitions only}.
	
Let $\mathcal{S} = \mathcal{Y} \times \mathcal{Y}$ represent the set of sensitive transitions, i.e.,
$(y_{i-1},y_i) \in \mathcal{S}$ if $p_{Y_i | X_i, Y_{i-1}}(y_i \given x_i,y_{i-1})$ is a function of $x_i$.
Moreover define
\begin{equation}
	\phi(p) =\left\{ \begin{array}{cl} 0, & p = 0\\ p \log p, & p \neq 0 . \end{array} \right.
\end{equation}
Then
\begin{align}
	\lefteqn{\mathcal{I}(X;Y) =} & \\
	\nonumber
	& \sum_{x \in \mathcal{X}} p_X(x) \sum_{(y_{i-1},y_i) \in \mathcal{S}}
	\pi_{y_{i-1}} \phi(p_{Y_i | X_i, Y_{i-1}}(y_i \given x_i,y_{i-1}))\\
	\nonumber
	&-\sum_{(y_{i-1},y_i) \in \mathcal{S}}
	\pi_{y_{i-1}} \phi\left(\sum_{x \in \mathcal{X}} p_X(x) p_{Y_i | X_i, Y_{i-1}}(y_i \given x_i,y_{i-1})\right).
\end{align}
Using (\ref{eqn:Markov-last}), the matrix $Q$ for the desired receptor, and an appropriately selected $\Delta t$, we can calculate $\mathcal{I}(X;Y)$.


Since $Z^n$ is a hidden Markov process, calculating the IID capacity $\mathcal{I}(X;Z)$ from inputs to ion channel state is done in one of two ways: either using Monte Carlo techniques to evaluate the expectation in (\ref{eqn:MutualInformation0}), replacing $y^n$ with $z^n$; or finding upper and
lower bounds, generalizing the technique from \cite{ThomasEckford2015IEEEtrans_submission_arXiv}. In either case,
the probability of the hidden Markov process $z^n$ is obtained using the sum-product algorithm \cite{KschischangFreyLoeliger2001IEEE}. By the data processing inequality. $\mathcal{I}(X;Y) \geq \mathcal{I}(X;Z)$.

\section{Results}

\begin{figure}
	\begin{center}
	\includegraphics[width=3.28in]{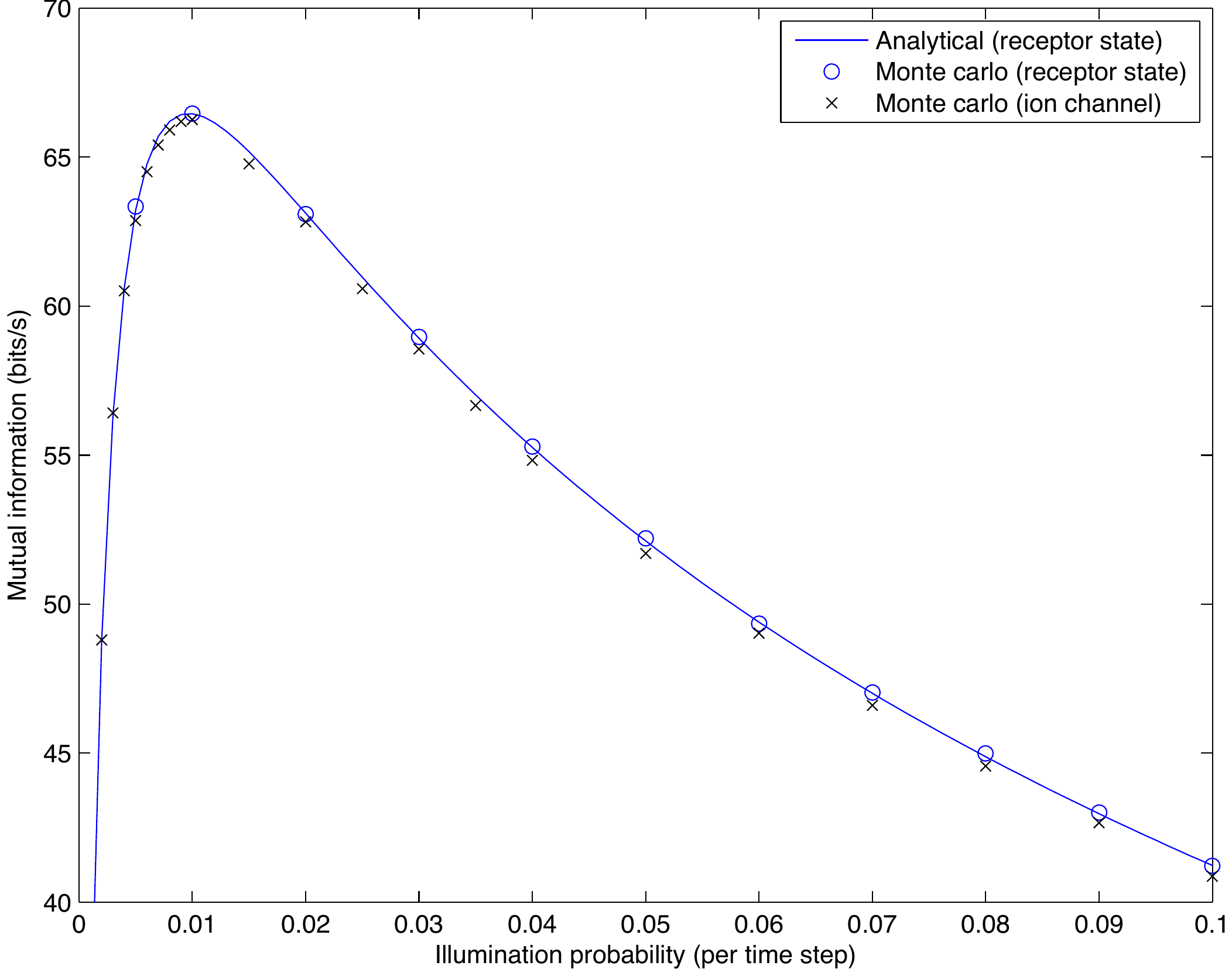}
	\includegraphics[width=3.28in]{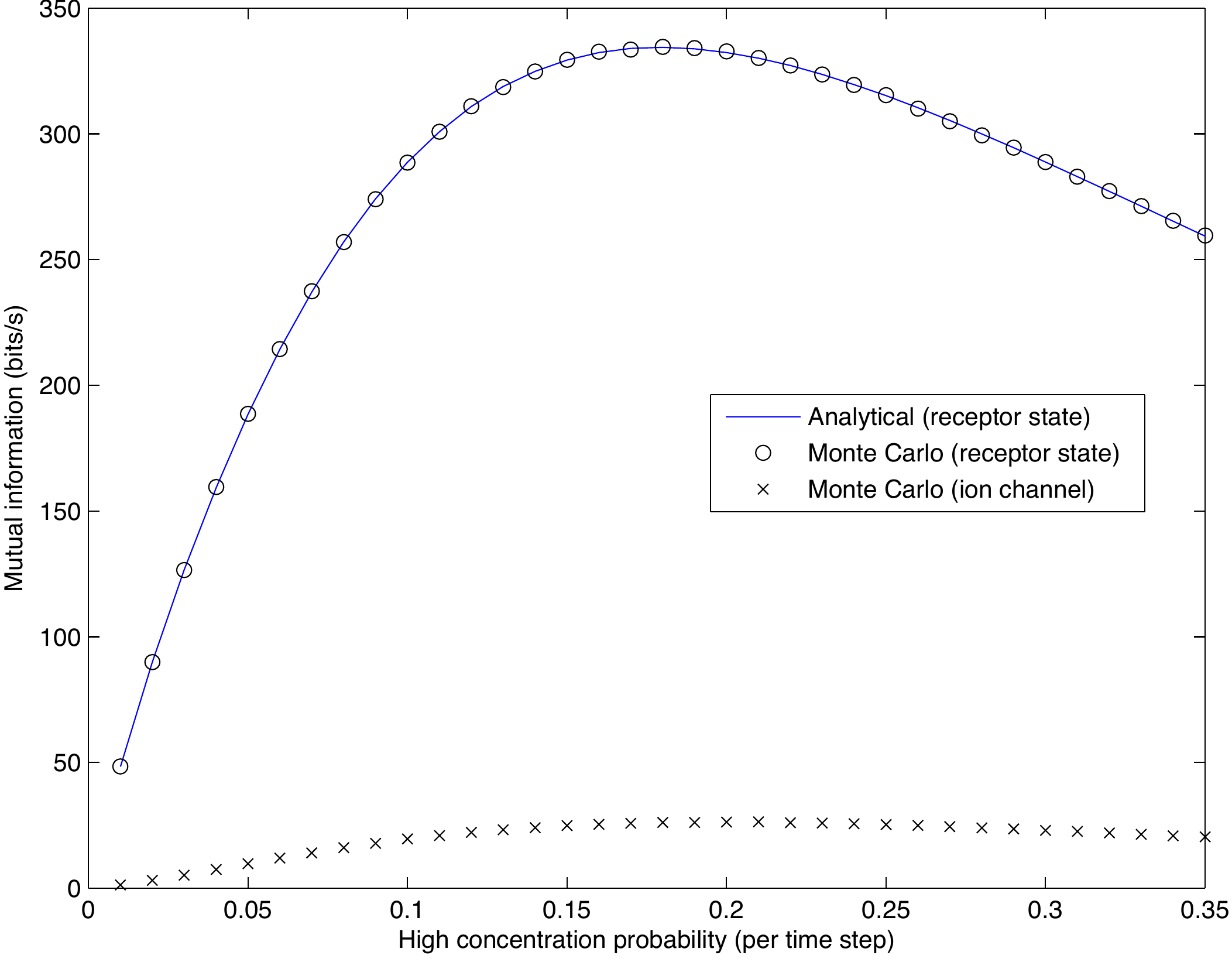}
	\end{center}
	\caption{\label{fig:AChResult} Mutual information with IID inputs for ChR2 (top figure) and ACh (bottom figure). Results for {\em receptor state} refer to $\mathcal{I}(X;Y)$, mutual information from input to the state of the receptor; results for {\em ion channel} refer to $\mathcal{I}(X;Z)$, mutual information from input to the state of the ion channel. }
\end{figure}

Mutual information results are given in Figure \ref{fig:AChResult}. The IID capacity may be found by taking the maximum of each curve. We see that when sensitive transitions are directly observable, there is a small gap between $\mathcal{I}(X;Y)$ and $\mathcal{I}(X;Z)$ (\textit{cf.}~ChR2); and when sensitive transitions are \emph{not} directly observable, there is a large gap (\textit{cf.}~ACh); thus, the receptor capacity is not always a tight bound for $\mathcal{I}(X;Z)$.  Heuristically, this gap appears to occur because of the structure of the channel.  The sensitive transition for ChR has stoichiometry $\mathbf{v}_{\text{ctrl}}=[-1,1,0]$ and the observation vector is $\mathbf{g}=[0,1,0]$; their inner product $\mathbf{v}_\text{crtl}\cdot\mathbf{g}^\text{T}=1$.  
In contrast, the three sensitive transitions for ACh have stoichiometries $\mathbf{v}^1_\text{ctrl}=[-1, 1, 0, 0, 0], \mathbf{v}^2_\text{ctrl}=[ 0, 0, 1, -1, 0], \mathbf{v}^3_\text{ctrl}=[ 0, 0, 0,1,-1]$ respectively; the observation vector is $\mathbf{g}=[1,1,0,0,0]$, and $\mathbf{v}^i_\text{crtl}\cdot\mathbf{g}^\text{T}=0$ for each $i$. 

Ideally, information theoretic analysis would lead to predictions comparable with experimental data.  However, receptor binding is part of a multistage channel that includes secretion and diffusion.  Channelrhodopsin is part of a multistage channel too: light-triggered currents can promote or inhibit action potentials depending on the type of ion coupled to the channel.  Both channels involve nonlinearities and memory effects that call for additional analysis.


\bibliographystyle{IEEEbib} 
\bibliography{MolecularInfoTheory,infotheory,signaling,Cowan,PJT,neuroscience,Dicty}

\end{document}